\documentclass[aps,pra,showpacs,amsmath,twocolumn,superscriptaddress]{revtex4-1}
\usepackage{amsmath,amssymb,graphicx}
\usepackage[final]{hyperref}
\usepackage{booktabs}

\begin{document}

\title{Effect of atomic diffusion on the Raman-Ramsey CPT resonances}

\author{Elena Kuchina}
\affiliation{Thomas Nelson Community College, Hampton VA 23666, USA}
\author{Eugeniy E. Mikhailov}
\author{Irina Novikova}
\email[inovikova@physics.wm.edy]{}
\affiliation{Department of Physics, College of William $\&$ Mary, Williamsburg, Virginia 23187, USA}

\date{\today}

\begin{abstract}
We experimentally investigated the characteristics of two-photon transmission resonances in Rb vapor cells with different amount of buffer gas  under the conditions of steady-state coherent population trapping (CPT)  and  pulsed Raman-Ramsey (RR-) CPT interrogation scheme. We particularly focused on the influence of the Rb atoms diffusing in and out of the laser beam. We showed that this effect modifies the shape of both CPT and Raman-Ramsey resonances, as well as their projected performance for CPT clock applications. In particular we found that at moderate buffer gas pressures RR-CPT did not improved the projected atomic clock stability compare to the regular steady-state CPT resonance.
\end{abstract}
		

\maketitle

\section{Introduction}

Precise measurements of the energy level splittings in atoms and molecules are in hearts of many devices. Long-lived transitions between the ground-state hyperfine sublevels of alkali atoms are particularly attractive for practical purposes due to their high quality factor. Since their transition frequencies are typically in a few GHz range, there has been a lot of interest in investigating all-optical interrogation methods for development of compact frequency standards and  magnetometers~\cite{vanier_book}.

\begin{figure}[htp]
  \includegraphics[width=1.0\columnwidth]{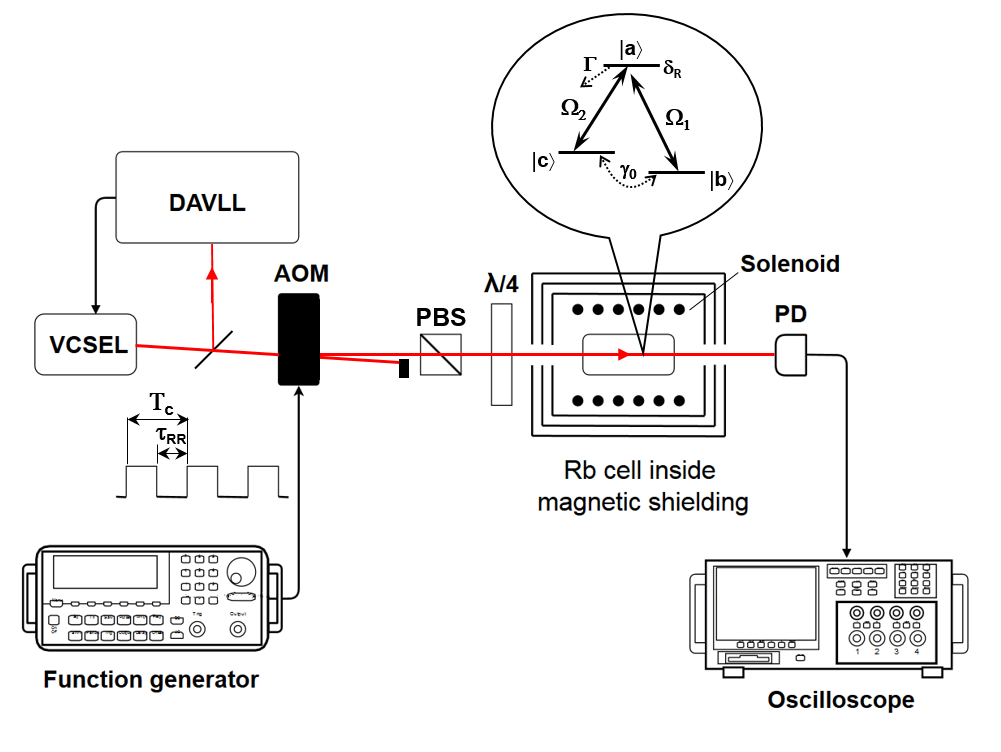}
  \caption{
  	\emph{(Color online)}
	Schematic of the experimental setup. The power and polarization of the optical field before the Rb vapor cell is controlled using an acousto-optical modulator (AOM), a polarizing beam splitter (PBS) and a quarter wave plate ($\lambda/4$). To stabilize the frequency of the VCSEL, approximately $20\%$ of its output is reflected into the DAVLL block, consisting of an auxiliary vapor cell in strong magnetic field, a quarter wave-plate and a balanced photodetector~\cite{yashchukRSI00}. Inset shows the simplified level structure of Rb atoms.
	\label{fig:setup}
	}
	
\end{figure}

Coherent population trapping (CPT)~\cite{arimondo'96po} is a two-photon optical effect, in which two optical fields form a resonant $\Lambda$-system based on the two ground hyperfine states, as shown in Fig.~\ref{fig:setup}. Under the two-photon resonance conditions $ \omega_{bc} + \nu_1 - \nu_2$, i.e., when the difference of the two optical field frequencies $\nu_{1,2}$ matches the energy splitting of the two ground states $\omega_{bc}$, the combined action of the two optical fields optically pumps the atoms into a non-interacting coherent superposition of these two states known as a ``dark state'' $|D\rangle$:
\begin{equation}
|D\rangle = \mathcal{N} \left(\Omega_1 |c\rangle - \Omega_2 |b\rangle \right),
\end{equation}
where $\Omega_{1,2}$ are the Rabi frequencies of the two applied optical fields, and $\mathcal{N}= 1/\sqrt{\Omega_1^2 + \Omega_2^2}$ is the normalization constant. Under the CPT conditions the absorption of the atomic medium is suppressed, and one observes a narrow transmission resonance with width $\gamma_\mathrm{CPT}$:
\begin{equation}
\gamma_\mathrm{CPT} = \gamma_0 + |\Omega|^2/\Gamma,
\end{equation}
where the first term represents  the dark state decoherence rate $\gamma_0$, and the second term demonstrates the CPT resonance power broadening, as $|\Omega|^2 = |\Omega_1|^2 + |\Omega_2|^2 $, and $\Gamma$ is an effective excited state decay rate~\cite{javan'02}.
In a dilute alkali metal vapor the decoherence rates of the ground-state sublevels are often governed by the thermal motion of the atoms, leading to relatively long dark state life time: a few hundreds microsecond for a vapor cell with a buffer gas, and up to hundreds of milliseconds for the cell with anti-relaxation wall coating~\cite{wynands'99,XiaoMPL09,novikovaLPR12}. 

In the last few decades CPT resonances have been successfully implemented for both compact and chip-scale atomic clocks and 
magnetometers~\cite{vanier05apb,knappeCM07}.  A traditional scheme for a CPT-based frequency standard operation~\cite{vanier05apb} includes a phase-modulated laser (typically VCSEL), that produces necessary optical fields interacting with atoms in a desired $\Lambda$ configuration. If the phase modulation frequency is adjusted precisely to the frequency difference between the two atomic ground levels, the optical transmission peaks, allowing the frequency of the laser modulation source to be locked to the atomic frequency. 
For atomic clocks the external source is typically locked to the magnetic-field insensitive atomic transition, whereas for CPT-based magnetometers the frequency difference between two magnetic-sensitive transitions is measured to extract information about magnetic field value. 

There is an on-going effort to improve the characteristics of CPT resonances, for example to reduce their sensitivity to various technical noises (laser intensity and frequency noise, fluctuations of environmental conditions), to increase their contrast, etc. 
It has been demonstrated, for example, that by tracking the time evolution of the dark state (similar to the more traditional Ramsey scheme with two separated interrogation zones~\cite{PhysRevLett.48.867}) offer several important advantages over traditional steady-state CPT transmission measurements. This so-called Raman-Ramsey interaction~\cite{ZanonPRL05} works as followed: first, the bi-chromatic optical field is turned on for long enough time to prepare atoms in the dark state; then, the optical field is turned off for the time $\tau_{RR}$, and the dark state is allowed to freely evolve in the dark; finally, the optical fields are turned on again, and their transmission after the time $\tau_m$ is monitored. In case of non-zero two-photon detuning, the dark state is not stationary, but it acquires an extra relative phase during its time evolution:
\begin{equation}
|D(\delta_R)\rangle = \mathcal{N} \left(\Omega_1 |c\rangle - \Omega_2 e^{i \delta_R \cdot \tau_{RR}} |b\rangle \right),
\end{equation}
where $\delta_R = \omega_{bc} - \nu_1 + \nu_2$ is the two-photon Raman detuning.
It is easy to see now that after the time $\tau_{RR} = \pi/\delta_R$ the original dark state evolves into the strongly-interacting bright state, and if the relative phases of the two optical fields are maintained, instead of enhanced transmission, one observes enhanced absorption. It is also clear that the oscillations between extra transmission and absorption should  be a periodic function of the evolution time $\tau_{RR}$, as long as this time is shorter than the ground-state coherence life-time. Assuming the homogeneity of the ground-state decoherence (i.e., that all atoms experience the same ground-state decay rate $\gamma_0$), it is possible to calculate the expected RR-CPT absorption coefficient $\kappa_{RR}$ ~\cite{zanonPRA2011}:
\begin{equation}
\kappa_{RR} = \alpha \left( 1 + \beta e^{-\gamma_0\tau_{RR}}cos(\delta_R\tau_{RR}-\Phi)\right), \label{RRCPTshape}
\end{equation}
where the values of the coefficients $\alpha, \beta$ and $\Phi$ are calculated in Ref.~\cite{zanonPRA2011}.

Several publications theoretically and experimentally demonstrated the advantages of the Raman-Ramsey interrogation method compared to the traditional cw CPT~\cite{zanonIEEE05,ClaironIEEE09,MitsunagaPRA13,Liu:13,kitchingPRA15}. In addition to typically having a larger contrast, an attractive feature of Raman-Ramsey CPT resonance (RR-CPT) is that the width of the observed resonances is determined only by the evolution time, and does not depend on laser power, unlike the regular CPT resonance. Thus, such resonances are not susceptible to the power broadening and light shifts, as was demonstrated in several publications~\cite{ClaironIEEE09,PatiJOSAB15,kitchingPRA15}. 
In particular, one can show that for a homogeneously power-broadened CPT resonance, in which all atoms have the same ground-state decoherence rate, the expected enhancement in the signal-to-noise ratio is:
\begin{equation} \label{RR-CPT-fom}
\frac{\mathrm{SNR}_{RR}}{\mathrm{SNR}_{CPT}} = 2 (\pi \tau_{RR} \gamma_0)^2 \frac{C_{RR}}{C_{CPT}} \sqrt{\frac{\tau_m I_{RR}}{T_c I_{CPT}}},
\end{equation}
where $T_c$ is the total duration of one Ramsey pulse sequence, $C_{RR}$ and $C_{CPT}$ are the contrasts of the Raman-Ramsey and traditional CPT resonances, correspondingly, and $I_{RR}$ and $I_{CPT}$ are the corresponding average background intensities. It is easy to estimate, that the RR-CPT arrangement is the most advantageous for $\tau_{RR} \sim 1/\gamma_0$.

In this paper, we consider the case of ``inhomogeneously'' broadened ground state coherence, in which different atoms can have substantially different decoherence rate $\Gamma_0$. This situation often occurs in atomic vapor cells with buffer gas, in which atoms can diffuse out of the interaction volume, and then come back after spending some time outside of the laser beams without dephasing its ground-state coherence. Such diffusion process introduces atoms with wide range of lifetimes inside the interaction region, causing strong modifications in the CPT lineshape, in particular, the appearance of a sharp ``pointy'' top~\cite{novikovaJMO05,xiaoPRL06}. It was shown that such behavior can be qualitatively described similarly to the Raman-Ramsey resonances~\cite{xiaoPRL06,XiaOE08, XiaoMPL09}, but by averaging over the possible values of the evolution in the dark time, determined by dynamics of the atomic diffusion. 

Below, we experimentally study this different regime for Raman-Ramsey CPT effect, in which an atom undergo both controlled and random evolution in the dark: one due to turning the light fields on and off, and the other due to the  atom temporarily leaving the interaction region. We compare the spectral lineshapes of regular CPT resonances (continuous laser interrogation) and Raman-Ramsey CPT resonances (pulsed interrogation) in two different Rb vapor cells with different amount of buffer gas ($5$~Torr and $30$~Torr of Ne) and observe a significant modification of the RR-CPT lineshape from the Eq.~\ref{RRCPTshape}.
We present a brief comparison of the expected sensitivity for the frequency measurements using steady-state CPT and RR-CPT resonances for the Rb vapor cells with different buffer gas pressure, and demonstrate that the diffusion-induced lineshape modifications change the relative figure of merit between the two interrogation methods. 

\section{Experimental Arrangements}

In our experiment, we used a temperature-stabilized vertical  cavity   surface-emitting  diode  laser  (VCSEL) operating at the Rb D$_{1}$ line ($\lambda=795$~nm).  The laser was  current-modulated  at $\nu_{rf} = 6.8347$~GHz  such that the laser carrier frequency and the first  modulation  sideband were tuned to the  $5S_{1/2}F=2  \rightarrow 5P_{1/2}F'=1$  and  $5S_{1/2}F=1  \rightarrow 5P_{1/2}F'=1$ transitions of ${}^{87}$Rb correspondingly. The two-photon Raman detuning $\delta_R$ was controlled by adjusting the laser microwave modulation frequency $\nu_{rf}$ around the value of the ${}^{87}$Rb hyperfine splitting using  a computer-controlled home-made microwave source ~\cite{mikhailov2010JOSAB_linparlin_clock}, such that $\delta_R=\nu_{rf}-6.834687135$~GHz. The intensity ratio
between  the  sideband  and  the  carrier  was  adjusted  by  changing the modulation  power  sent   to  the  VCSEL, and  kept the sideband to  carrier ratio equal to  60\%. The optical frequency of the laser was stabilized using  a   dichroic-atomic-vapor  laser lock  (DAVLL)~\cite{yashchukRSI00}. The  details   of  the construction  and   operation  of the home-made laser system  are  provided  in
\cite{mikhailov2009AJP_clock_for_undergrads}. 

The circularly-polarized laser beam  with maximum total  power 90~$\mu$W and a  slightly elliptical Gaussian  profile  [1.8~mm and  1.4~mm  full  width half  maximum  (FWHM)] traversed a cylindrical Pyrex cell (length 75~mm; diameter 22~mm) containing isotopically enriched ${87}$Rb vapor and either 5~Torr or 30~Torr of Ne buffer gas. The cell was mounted inside a three-layer magnetic shielding and actively temperature-stabilized at $53^\circ$C. To isolate the magnetic field-insensitive CPT resonance, we applied a homogeneous longitudinal magnetic field of $520$~mG using a solenoid mounted inside the innermost magnetic  shield. 
Changes in the total laser transmission were recorded using a photodetector (PD), placed after the Rb cell. 

For the Raman-Ramsey CPT measurements, we turned the laser beam on and off using an acousto-optical modulator (AOM), placed before the vapor cell, modulated with a square wave at the frequency $f_{mod}$. We verified that each ``on'' half-cycle was long enough to achieve the steady state CPT conditions. The following half-cycle, which corresponded to the AOM ``off'' period, served as the dark evolution time $\tau_{RR} = (2f_{mod})^{-1}$. To reproduce the Raman-Ramsey signal, shown in Fig. 2(c-f) the values of the laser transmission were recorded $\tau_m=20~\mu$s after the laser was turned on again. 

\section{Experimental Results}

\begin{figure}[ht]
\centering
\fbox{\includegraphics[width=\linewidth]{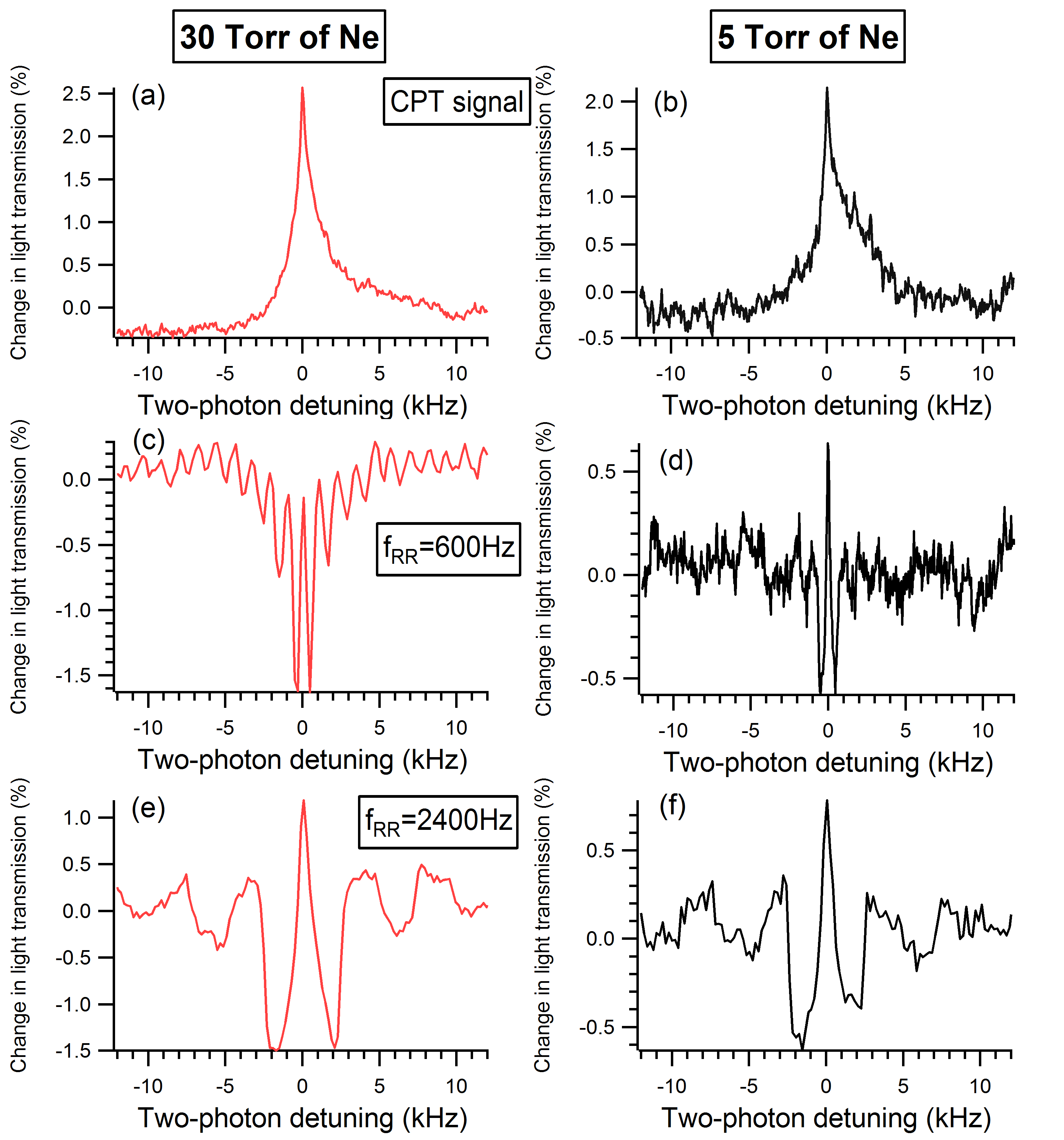}}
\caption{Examples of steady-state CPT resonances (top row) and Raman-Ramsey resonances obtained for different modulation frequencies $f_{mod}$ and, hence, for different dark time intervals (two bottom rows). The left column corresponds to the signals obtained in the Rb cell with $30$~Torr of Ne buffer gas, and the right one - in the Rb cell with $50$~Torr of Ne buffer gas. In all graphs the vertical axis represents the relative optical transmission ($I(\delta_R)/I_{background} - 1$. }
\label{fig:examplespectra}
\end{figure}

Fig.~\ref{fig:examplespectra}(a,b) presents sample spectra of steady-state CPT resonances recorded in the Rb vapor cells with different amount of the buffer gas under otherwise identical experimental conditions. For atoms traversing the illuminated interaction region only once, the dark state lifetime can be estimated from the average diffusion time through the laser beam of radius $a$~\cite{happer'72}, implying the decoherence rate to be:
\begin{equation}
\gamma_0 \simeq (2.405)^2 D_0 \frac{p_0}{p} \frac{1}{a^2},
\end{equation}
where $D_0 = 0.2~\mathrm{cm}^2/s$ is the diffusion constant of Rb atoms in Ne at the atmospheric pressure $p_0 = 760$~Torr, and $p$ is the buffer gas pressure inside the cell. Also, since in this case all atoms experience approximately  the same ground-state decoherence rate, the CPT lineshape is expected to be Lorentzian (this is the necessary assumption for deriving Eq.~\ref{RR-CPT-fom}). In this model the expected minimum full width half maximum (FWHM) values for the CPT resonance in the cells with $5$~Torr and $30$~Torr of Ne buffer gas are $5.6$~kHz and $930$~Hz correspondingly. In our experiments, however, the measured FWHM values for the CPT resonances were quite similar in both cells: $\approx 1300$~Hz and $\approx 1100$~Hz. Also, the shapes of the resonances were clearly non-Lorentzian, indicating the contributions from atoms with a wide range of the ground state decoherence rates.

The pointy lineshape analysis of these resonances reveals a strong influence of atomic diffusion in and out of the laser beam~\cite{novikovaJMO05,xiaoPRL06,XiaOE08,XiaoMPL09}. To understand the origin of the CPT linewidth narrowing, one must take into account the possibility of an atom to leave the interaction region and then return after some time without dephasing its quantum state. Such repeated interaction model, developed in Ref.~\cite{xiaoPRL06,XiaOE08}, draws on parallels between the atom multiple interactions with the laser beams interspersed with the ``evolution in the dark'' while the atom is outside of the interaction region with the Raman-Ramsey CPT resonances where the time of the evolution in the dark is controlled by turning on and off the laser fields. However, since now the evolution time is governed by the diffusion dynamics and thus stochastic, the resulting signal is effectively averaged over the distribution of the evolution times, which interfere constructively only for $\delta_R=0$, leading to a ``peaky'' CPT resonance. 

Next, we analyze the effect of such diffusive atomic motion on the Raman-Ramsey CPT resonances. The examples of RR-CPT fringes recorded in both vapor cells at two different AOM modulation frequencies are shown in Fig.\ref{fig:examplespectra}(c-f). For the $f_{mod}=600$~Hz ($\tau_{RR} = 830~\mu$s), the recorded transmission signals resemble the traditional RR-CPT fringes, for which the dark evolution time is comparable to the dark state decoherence time: a few higher-contrast fringes near $\delta_R=0$, and the reduced contrast fringes for larger Raman detunings. Interestingly, the relative heights of the central fringe in the two cells are almost the same, even though one might not have expected to see any fringes for such long dark time for the cell with $5$~Torr of Ne, since the diffusion time of Rb atoms through the laser beams is only $\approx 60~\mu$s. Thus, the observed RR-fringes are mainly due to the atoms that diffuse back to the interaction region after $\tau_{RR}$ time. 

In the RR-CPT spectra recorded for the shorter dark evolution time (modulation frequency $f_{mod}=2400$~Hz, $\tau_{RR} = 210~\mu$s) the shape of the fringes changes: there is a clear sharp peak at zero Raman detuning, similar to the sharp peak observed for CPT resonances, with some additional sharp dispersion-like structures for $\delta_R$ equal to the multiple of $f_{mod}$. The shape of the central fringe implies significant contribution from the atoms with the diffusion times longer than $\tau_{RR}$. Since these atoms experience similar distributions of dark evolution times, and their sensitivity to small changes of $\delta_R$ is similar for both steady-state CPT or RR-CPT detection schemes. 

\begin{figure}[htbp]
\centering
\fbox{\includegraphics[width=\linewidth]{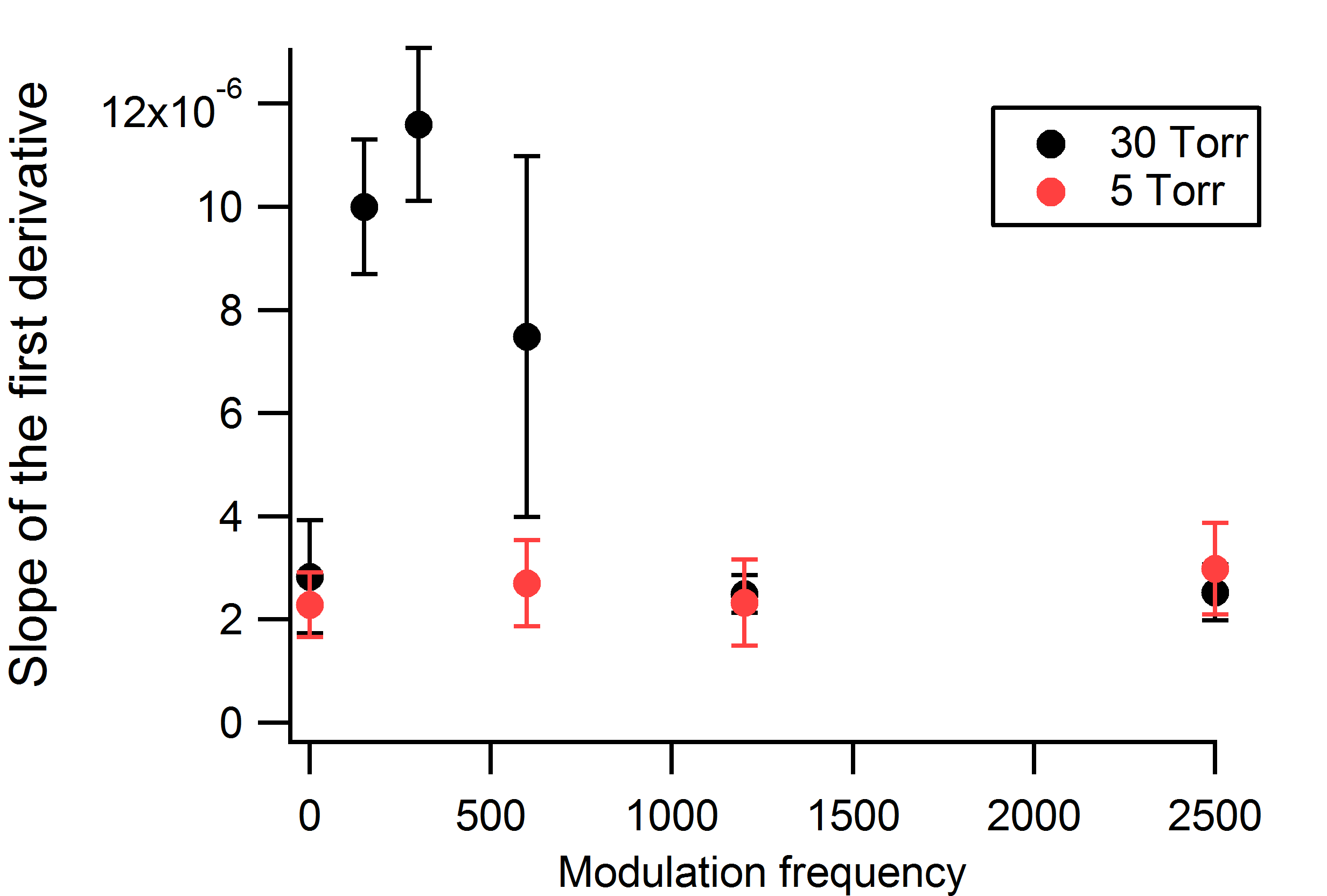}}
\caption{Estimated value of the error signal slope for atomic clock locking based on the analysis of the measured transmission resonances. Data points corresponding to zero modulation frequency correspond to the steady-state CPT resonances. Each data point is the result of averaging over 5-10 RR-CPT spectra, and  error bars represent one standard deviation spread.}
\label{fig:slopes}
\end{figure}

To verify this statement and to estimate the projected performance of a possible frequency standard, we calculate the maximum slope of the first derivative of the laser transmission $(\partial I/ \partial \delta_R)/I|_{\delta_R=0}$, where $I$ is the measured transmitted intensity.  Since the background transmission values in the both cells were similar, the value of this slope can be used to estimate the error signal for the feedback loop when locking the microwave modulation source to a CPT or RR-CPT resonance. 
The values of measured slopes for the two vapor cells are shown in Fig.~\ref{fig:slopes}. 
It is clear that RR-CPT resonances in the $5$~Torr vapor cell do not provide any improvement in performance compare to the standard steady-state CPT arrangement, as the slope values are independent of the Raman-Ramsey modulation frequency $f_m$. Indeed, in this regime the longest dark state evolution time is provided by the atom physically diffusing out of the laser beam and coming back, that contribute similarly in the sharpness of the central spectral feature for both CPT and RR-CPT resonances. Thus, having longer or shorter $\tau_{RR}$  only affects of the number of probed atoms, without changing the resonance linewidth. 

Our measurements in the $30$~Torr cell, however, indicated that for lower modulation frequencies there is a clear advantage of using RR-CPT detection method, as we observed almost 6-fold increase of the error signal for $f_m=300$~Hz. The higher buffer gas pressure slows the diffusion dynamics, and it takes much longer for atoms to leave the interaction region. Still, for shorter dark evolution times $\tau_{RR}$ (i.e. for higher modulation frequencies $f_m > 1000$~Hz) the measured maximum slopes of the laser transmission near $\delta_R = 0$ again become comparable to those of the regular CPT resonances, indicating the return to the regime in which the highest spectral sensitivity is determined by the diffusing atoms reentering the laser beam. 

\section{Conclusions}

We analyzed the Raman-Ramsey CPT spectra, recorded in the Rb vapor cells with $5$ and $30$~Torr of Ne buffer gas in the regime where the diffusion of the atoms in and out of the illuminated interaction region played significant role. Previous studies have shown the superiority of the RR-CPT detection compare to the traditional CPT transmission resonances for precise frequency detection in the case of homogeneous ground-state atomic decoherence. In the conditions of our experiments, however, we found that the  ratio between the evolution in the dark time of the Raman-Ramsey sequence $\tau_{RR}$ and the average dark state evolution time of diffusing atoms becomes an important parameter for evaluating the advantages of RR-CPT method. In particular, we found that in case of shorter $\tau_{RR}$ the most sensitive frequency response to the changes of the two-photon Raman detuning is provided by the atoms with long diffusion times, and thus using a more complicated RR-CPT method does not lead to any increase in sensitivity. 

\section{Acknowledgments}

We acknowledge the support of National Science Foundation Women in Scientific Education (WISE) Award HRD‐1107147 for making this collaboration possible. We also would like to thank Melissa Guidry for her help at the early steps of the Raman-Ramsey setup construction and Owen Wolfe for assistance with the data analysis. 



\end{document}